\documentclass[a4paper]{jpconf}
\usepackage[dvips]{graphicx}
\usepackage{amssymb}
\usepackage{iopams}

\begin{document}
\title{Phase separation in high-T$_c$ cuprates}

\author{A S Moskvin$^{1,2}$, Yu D Panov$^1$}

\address{$^1$ Institute of Natural Sciences and Mathematics, Ural Federal University, 620083 Ekaterinburg, Russia
}
\address{$^2$ Institute of Metal Physics UB RAS, 620108 Ekaterinburg, Russia
}
\ead{alexander.moskvin@urfu.ru}

\begin{abstract}
We develop a minimal non-BCS model for the CuO$_2$ planes  with the on-site Hilbert space  reduced to only three effective valence centers CuO$_4$ with different charge, conventional spin, and orbital symmetry, combined in a charge triplet, to describe the low-energy electron structure and the phase states of HTSC cuprates. Using the $S$\,=\,1 pseudospin algebra we introduce an effective spin-pseudospin Hamiltonian which takes into account local and nonlocal correlations, one- and two-particle transport, and spin exchange. To illustrate the possibilities of the molecular field approximation we start with the analysis of the atomic and the "large negative-$U$" limits of the model in comparison with the Bethe cluster approximation, classical and quantum Monte Carlo methods. Both limiting systems exhibit the phase separation effect typical of systems with competing order parameters.
 The $T$\,-\,$n$ phase diagrams of the complete spin-pseudospin model were reproduced  by means of a site-dependent variational approach within effective field approximation typical for spin-magnetic systems. Limiting ourselves to two-sublattice approximation and $nn$-couplings we arrived at several N\'{e}el-like phases in CuO$_2$ planes for parent and doped systems with a single nonzero local order parameter: antiferromagnetic insulator, charge order, glueless $d$-wave Bose superfluid phase, and unusual metallic phase. However, the global minimum of free energy is realized for phase separated states which  are bounded by the third-order phase transition line $T^{\star}(n)$, which is believed to be responsible for the onset of the pseudogap phenomenon. With a certain choice of the Hamiltonian parameters the model phase diagrams can quite reasonably reproduce the main features of experimental phase diagrams for T- and T$^{\prime}$-cuprates and novel nickelates. The superconducting phase of cuprates/nickelates is determined by the on-site composite boson transport, it is not a consequence of pairing of doped holes/electrons, but represents one of the possible phase states of parent systems.
\end{abstract}

\section{Introduction}

Despite the past 35 years since the discovery of HTSC in cuprates\,\cite{HTSC}, the explanation of their unusual normal and superconducting properties still remains a challenge for the condensed matter community. In particular, the most studied hole-doped T-type cuprates are characterized by uniquely complex phase diagrams with coexistence of a large variety of phase states and specific temperature regimes. Many researchers argue that the main features of the cuprates superconductors phase diagram can be derived considering the disorder and intrinsic mesoscopic static/dynamic phase separation as a key property of these materials (see, e.g. Refs.\,\cite{deMello,Pelc,Liarocapis,diCastro}). However, a correct description of a phase-inhomogeneous state presupposes the use of an adequate physical model. Recent discoveries of anomalous properties of cuprates and nickelates with a T$^{\prime}$-structure\,\cite{Naito-2016,Li}, that is with no apical oxygen, including HTSC in parent compositions, firstly indicate the crucial role of apex oxygen, and, secondly, indicate the need to abandon the generally accepted concept of the parent composition as a Mott-Hubbard  antiferromagnetic insulator. Instead, we propose to introduce a more universal, albeit somewhat formalized, definition of the "parent" system of CuO$_2$/NiO$_2$ planes with a nominal 3d$^9$ configuration for Cu/Ni sites, or "half-filling", which, depending on the parameters determined by the "out-of-plane" potential and electron-lattice relaxation, can be found in various states, from an antiferromagnetic or non-magnetic insulator, Fermi metal, to a high-temperature superconductor. Furthermore, from our point of view, the explanation of the cuprate/nickelate puzzle should include some fundamentally new physics, which requires not only going beyond the Bardeen-Cooper-Schrieffer  (BCS) paradigm of the superconducting state\,\cite{Hirsch-2009}, but also revising a number of other well-established concepts.

Following the spin-magnetic analogy proposed by Rice and Sneddon\,\cite{Rice_1981} to describe the three charge states (Bi$^{3+}$, Bi$^{4+}$, Bi$^{5+}$) of the bismuth ion in BaBi$_{1-x}$Pb$_x$O$_3$, earlier we started to develop a  minimal "unparticle"\, model for the CuO$_2$ planes with the "on-site"\, Hilbert space of the CuO$_4$ plaquettes to be a key element of crystal and electron structure of high-T$_c$ cuprates,  reduced to states formed by only three effective valence centers [CuO$_4$]$^{7-,6-,5-}$ (nominally Cu$^{1+,2+,3+}$, respectively), forming a "well isolated"\, charge triplet\,\cite{truegap,dispro,Moskvin-JSNM-2019}. The very possibility of considering these centers on equal footing is predetermined by the strong effects of electron-lattice relaxation in cuprates\,\cite{Mallett,Moskvin-PSS-2020}.

Electrons of such many-electron atomic species with strong $p$\,-\,$d$ covalence and strong intra-center correlations cannot be described within any conventional (quasi)particle approach that addresses the [CuO$_4$]$^{7-,6-,5-}$    centers within the on-site
hole  representation $|n\rangle$, n = 0,\,1,\,2, respectively.
Instead of conventional quasiparticle $\bf k$-momentum description we make use of a real space on-site "unparticle"\, $S$\,=\,1 pseudospin formalism to describe the charge triplets and introduce an  effective spin-pseudospin Hamiltonian which takes into account both local and nonlocal correlations, single and two-particle transport, as well as Heisenberg spin exchange interaction.

In this paper we use the   charge triplet model  to reproduce the $T$\,-\,$n$ phase diagrams of the CuO$_2$ planes  within  a site-dependent effective field approximation, which treats the on-site quantum fluctuations exactly and all the intersite interactions within the mean-field approximation (MFA) typical for spin-magnetic systems\,\cite{Condmat-2021}. Our main goal is to demonstrate that similar to other systems with competing order parameters the phase separation (PS) is a key property of cuprates/nickelates which can explain main features of their phase diagrams.

\section{Effective spin-pseudospin Hamiltonian}
The S\,=\,1 spin algebra includes the eight independent nontrivial pseudospin operators, the three dipole and five quadrupole operators:
$$
	{\hat S}_z;\, {\hat S}_{\pm}=\frac{1}{\sqrt{2}}({\hat S}_{x}\pm i{\hat S}_{y});\,{\hat S}_z^2;\,{\hat T}_{\pm}=\{{\hat S}_z, {\hat S}_{\pm}\};\,{\hat S}^2_{\pm} \, .
$$
The ${\hat S}^2_{\pm}$ operators  are actually the creation/annihilation operators of the composite on-site hole boson\,\cite{Condmat-2021}. Simplified, this boson is a pair of holes coupled by local correlations both with each other and with the "core", that is, the electronic  center [CuO$_4$]$^{7-}$ ($M_S$\,=\,-1). In fact, such a local boson exists only as an indivisible part of the Zhang-Rice hole center [CuO$_4$]$^{5-}$ ($M_S$\,=\,+1). Obviously, the  mean value $\langle {\hat S}_{\pm}^{2} \rangle\,=\,\frac{1}{2}(\langle {\hat S}_x^2-{\hat S}_y^2\rangle \pm i\langle\{{\hat S}_x,{\hat S}_y\}\rangle )$ can be addressed to be a complex superconducting local order parameter with a $d$-wave symmetry\,\cite{truegap,Condmat-2021}.
In lieu of "single-particle" ${\hat S}_{\pm}$ and ${\hat T}_{\pm}$ operators one may use  two novel operators:$
	{\hat P}_{\pm}=\frac{1}{2}({\hat S}_{\pm}+{\hat T}_{\pm});\,{\hat N}_{\pm}=\frac{1}{2}({\hat S}_{\pm}-{\hat T}_{\pm})$,
which do realize transformations Cu$^{2+}$$\leftrightarrow$\,Cu$^{3+}$ and Cu$^{1+}$$\leftrightarrow$\,Cu$^{2+}$, respectively.
However, strictly speaking, we should extend the on-site Hilbert space to a spin-pseudospin quartet $|SM;s\nu\rangle$: $|1\pm 1;00\rangle$ and $|10;\frac{1}{2}\nu\rangle$,
where $\nu =\pm\frac{1}{2}$, and instead of spinless operators ${\hat P}_{\pm}$ and ${\hat N}_{\pm}$ introduce operators ${\hat P}_{\pm}^{\nu}$ and ${\hat N}_{\pm}^{\nu}$,
which transform both on-site charge (pseudospin) and spin states\,\cite{Condmat-2021}.
\begin{figure*}[t]
\begin{center}
\includegraphics[width=10cm,angle=0]{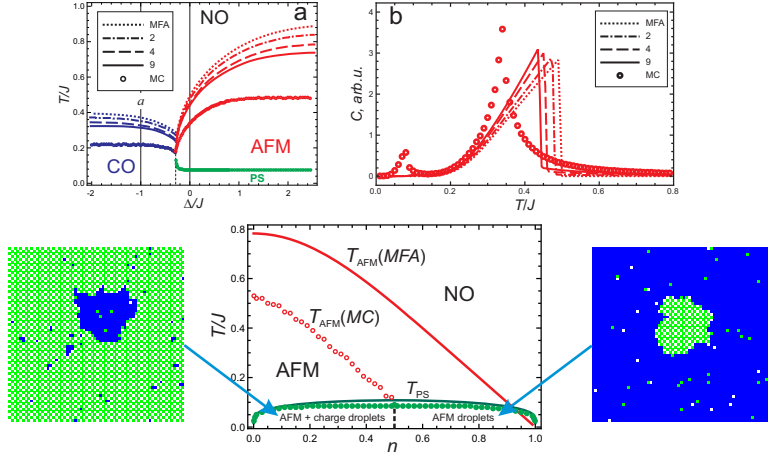}
\caption{(Color online) Top panel (a): $T$\,-\,$\Delta$ phase diagram for atomic limit given $n$\,=\,0.1,  $V$\,=\,0.1  as derived by MFA, Bethe cluster models and by the MC calculations.
Top panel (b): The temperature dependence of specific heat obtained by the MFA, Bethe cluster approximation and by the MC calculations.
  Bottom panel: $T$\,-\,$n$ phase diagram for strong exchange limit. Open circles denote the MC results for the maxima of susceptibility
due to the AFM ordering, and filled circles show the maxima of the specific heat
at the PS transition. Solid lines show the value of the calculated mean-field critical temperatures $T_{AFM}$ and $T_{PS}$\,\cite{Panov_JLTP_2017,Panov_JETPLett_2017}.}
\label{fig1}
\end{center}
\end{figure*}
As for conventional spin-magnetic systems, we can integrate out the high-energy degrees of freedom, and after projecting onto the Hilbert basis of well isolated charge triplet, or spin-pseudospin quartet, we have chosen, to arrive at the effective spin-pseudospin Hamiltonian obeying the spin and pseudospin kinematic rules with a charge density constraint, $\frac{1}{N}\sum _{i} \langle {\hat S}_{iz}\rangle =n$, where $n$ is the deviation from a half-filling:
\begin{equation}
{\hat H}={\hat H}_{pot}+{\hat H}_{kin}^{(1)}+{\hat H}_{kin}^{(2)}+{\hat H}_{ex}\,;
\label{H}	
\end{equation}
$$
	{\hat H}_{pot} =  \sum_{i}  (\Delta _{i}{\hat S}_{iz}^2
	  - \mu {\hat S}_{iz}) + \sum_{i>j} V_{ij}{\hat S}_{iz}{\hat S}_{jz}\,;
$$
$$
\fl{\hat H}_{kin}^{(1)}= -\sum_{i>j}\sum_{\nu} [t^p_{ij}{\hat P}_{i+}^{\nu}{\hat P}_{j-}^{\nu}+
 t^n_{ij}{\hat N}_{i+}^{\nu}{\hat N}_{j-}^{\nu}+
 \frac{1}{2} t^{pn}_{ij}({\hat P}_{i+}^{\nu}{\hat N}_{j-}^{\nu}+{\hat P}_{i-}^{\nu}{\hat N}_{j+}^{\nu}) +h.c.] \,;
$$
$$
  {\hat H}_{kin}^{(2)}=-\sum_{i>j} t_{ij}^b({\hat S}_{i+}^{2}{\hat S}_{j-}^{2}+{\hat S}_{i-}^{2}{\hat S}_{j+}^{2}), \,\,
{\hat H}_{ex}=\sum_{i>j} J_{ij}(\hat {\bf s}_i\cdot \hat {\bf s}_j)	= s^2\sum_{i>j}J_{ij} (\boldsymbol{\sigma}_i\cdot  \boldsymbol{\sigma}_j)\, .
$$
The first on-site term in ${\hat H}_{pot}$, resembling single-ion spin anisotropy, describes the effects of a bare pseudospin splitting and relates with the on-site density-density interactions, $\Delta$\,=\,$U$/2, $U$ being the local correlation parameter, or pair binding energy for composite boson. The second term   may be
related to a   pseudo-magnetic field $\parallel$\,$Z$ with $\mu$ being the hole chemical potential.  The third and fourth terms in ${\hat H}_{pot}$ describe the inter-site density-density interactions, or nonlocal correlations.
Kinetic energies ${\hat H}_{kin}^{(1)}$ and ${\hat H}_{kin}^{(2)}$ describe a single- and two-particle transport, respectively, where $t^p, t^n, t^{pn}$ and $t^b$ are  integrals for the correlated single-particle and the composite on-site boson hopping, respectively. Operator $\boldsymbol{\sigma}=2\hat{\rho}^s \mathbf{s}$ in ${\hat H}_{ex}$  takes into account the on-site spin density $\hat{\rho}^s=(1-{\hat S}_{z}^2)$. Depending on the values of the effective Hamiltonian parameters, the model predicts the possibility of realizing even for parent cuprates/nickelates both a typical antiferromagnetic insulating state (AFMI), charge order (CO), unusual Fermi liquid (FL) phase with electron-hole interplay and a bosonic superconductor (BS) phase with effective on-site hole bosons.
\begin{figure*}[t]
\begin{center}
\includegraphics[width=10cm,angle=0]{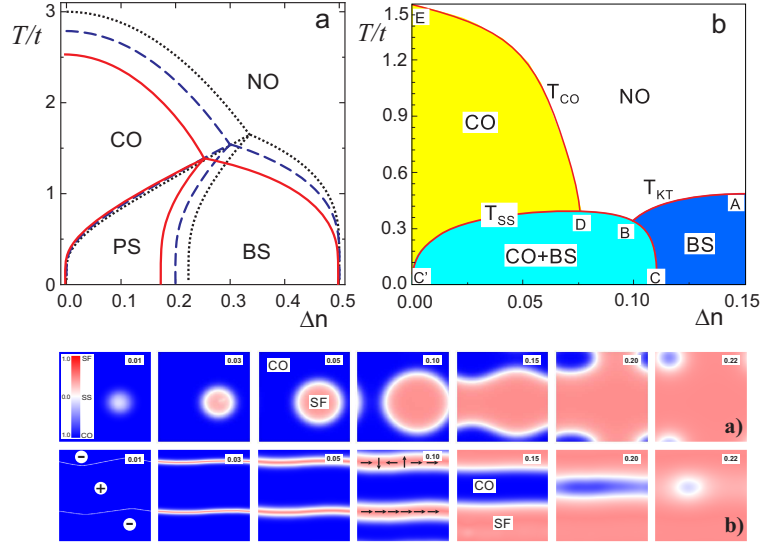}
\caption{(Color online) Top panels: T-$n$ phase diagram for 2D $hc$-bosons on square lattice at $V/t$\,=\,3. Panel (a): dotted line points to MFA data\,\cite{RMP}, dashed and solid lines indicate the results of the two- and four-site Bethe cluster approximation, respectively\,\cite{Panov}.  Panel (b): phase diagram for 2D hard-core bosons as derived by QMC\,\cite{Schmid2002}  technique.  A-B is the line of the Kosterlitz-Thouless phase transition, C-B-D-C$^{\prime}$ is the line of the first order phase transition, D-E  is the line of the second order Ising phase transition. Bottom panel: The two scenarios of the evolution of the $hc$-boson ground state configuration under doping away from half-filling. }
\label{fig2}
\end{center}
\end{figure*}

\section{Atomic and "large negative-$U$" limits}
The atomic limit of the spin-pseudospin model, which  assumes complete neglect for kinetic energy,  was considered within the MFA, Bethe cluster approach, and classical Monte-Carlo (MC) technique (see, e.g., Refs.\,\cite{Panov_JLTP_2017,Panov_JETPLett_2017}).   The $T$\,-\,$\Delta$ and  $T$\,-\,$n$ phase diagrams, the temperature dependence of specific heat, and the demonstration of the PS states for strong exchange limit are shown in Figure\,\ref{fig1}.  In the strong exchange limit both the Maxwell's construction technique and MC calculation point to a low-temperature "third-order" phase transition to the AFMI-CO PS regime with a distinct specific heat anomaly.

In the large "negative-$U$" approximation, which  assumes complete neglect for  single-particle kinetic energy and spin exchange, the CuO$_2$ plane becomes equivalent to hard-core ($hc$), or local boson system.
The conventional mean-field $T$\,-\,$n$ phase diagram for the hard-core bosons is well-known\,\cite{RMP}.
In addition to the conventional CO  and Bose superfluid BS phases the MFA predicts the appearance of an unconventional uniform supersolid phase (SS) with the "on-site"\, coexistence of the  insulating and  superconducting properties. However, detailed analysis of the $hc$ boson model shows that the SS phase is a mean-field artefact, de facto this homogeneous phase is intrinsically unstable. For $nn$-interactions and finite temperatures the "post-MFA" Maxwell's construction shows that the phase-separated CO-BS phase  has a lower energy than the uniform SS phase, though at $T$\,=\,0 the both phases have the same energy\,\cite{Panov}. Furthermore, this result is confirmed within Bethe cluster approximation as shown in the top panel (a) of the Figure\,\ref{fig2}.
 The bottom panel in the Figure\,\ref{fig2} demonstrates the two MC scenarios of the evolution of the $hc$-boson ground state configuration under doping away from half-filling with topologically different PS states.
For comparison in the top panel (b) of the Figure\,\ref{fig2} we present the phase diagram of the square lattice hard-core boson  model given  $V_{nn}=V=3t$, derived from the Quantum Monte-Carlo (QMC) calculations by Schmid {\em et al.}\,\cite{Schmid2002}.
At half-filling the system undergoes the charge ordering CO at $T_{CO}\approx $\,0.5\,$V$\,=\,1.5\,$t$\,\cite{Schmid2002} that is only slightly less than the exact Onsager answer for 2D Ising model ($t$\,=\,0):
However, under doping away from half filling, the checkerboard solid undergoes phase separation: the superfluid BS and solid CO phases coexist but not as a single thermodynamic supersolid phase as predicted in the mean-field approximation\,\cite{RMP}.
Thus, the phase separation turns out to be a typical phenomenon for systems with competing order parameters, described by particular versions of the model spin-pseudospin Hamiltonian.

\section{Effective-field approximation  for the complete spin-pseudospin model}
 \begin{figure*}[t]
\centering
\includegraphics[width=10 cm,angle=0]{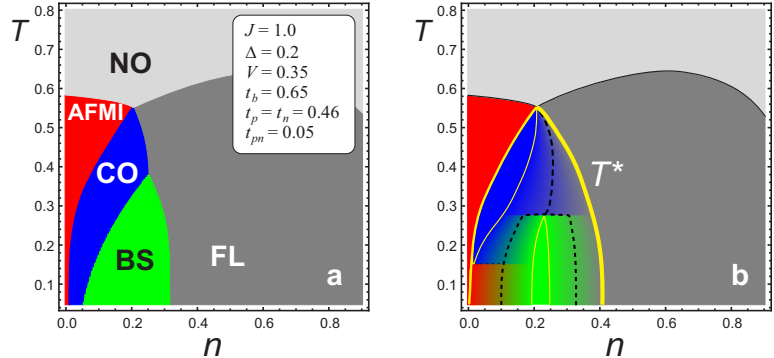}
\caption{(Color online) Model $T$\,-\,$n$ phase diagrams of the hole-doped cuprate calculated in the effective-field approximation
 ($n=p$ for the hole doping) with constant values of the Hamiltonian parameters (see inset); (a)  phase diagram assuming main homogeneous phases with no allowance made for the possible coexistence of two adjacent phases;  (b) phase diagram with the phase separation taken into account using Maxwell's construction. Dashed curves in (b) point to fifty-fifty volume fraction for two adjacent phases, yellow curves in (b) present the third-order phase transition lines, these limit areas with 100\% volume fraction. }
\label{fig3}
\end{figure*}
Making use of the effective field theory which combines the MFA with the exact accounting of local correlations, the variational approach (VA) based on the Bogolyubov inequality for the grand potential, the Caron-Pratt model for the on-site description of the FL phase\,\cite{Caron-Pratt}, we were able to numerically calculate the phase diagrams  of the complete spin-pseudospin system within the framework of a simplified model (two sublattices, nn-coupling,...)\,\cite{Condmat-2021}.
 The left hand side of the Figure\,\ref{fig3} shows an example of the phase diagram calculated given quite arbitrarily chosen parameters of the model Hamiltonian and assuming main homogeneous single-order parameter, or "monophases" with no allowance made for the possible coexistence of two adjacent phases. However, the numerical implementation of the Maxwell's construction shows that the minimum of free energy corresponds to PS realized in the region of coexistence of phases separated by the first-order phase transition lines. This works for phases AFMI-FL, AFMI-BS, CO-BS, CO-FL, and BS-FL, but not for AFMI-CO (see Figure\,\ref{fig3}, the right hand side panel).
  We can immediately note that the mysterious pseudogap phase is nothing more than the AFMI-FL-CO-BS PS state with the third-order phase transition line $T^*(n)$, which separates the gapless 100\% FL phase from the gapped phases, to be the highest pseudogap temperature.
  The PS model does predict several temperatures of the "third order"\, PS transitions limiting the PS phases, that is delineating areas with 100\% volume fraction,  and the temperatures of the percolation transitions, which can manifest itself in the peculiarities of the temperature behavior for different physical quantities.
\begin{figure*}[t]
\centering
\includegraphics[width=12 cm,angle=0]{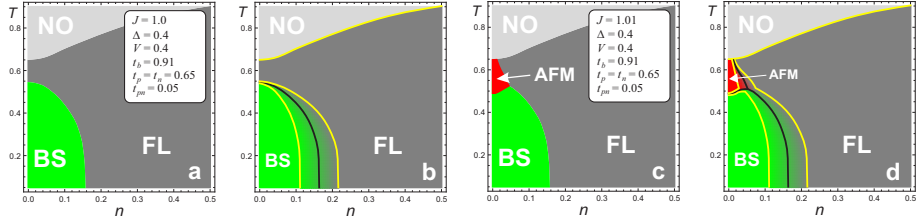}
\caption{(Color online) Model $T$\,-\,$n$ phase diagrams of the hole-doped cuprate/nickelate calculated in the effective-field approximation
  under constant values of the Hamiltonian parameters (see inset), which illustrate an AFMI-BS-FL and BS-FL interplay with (b,d) and without (a,c) PS taken into account.  Solid yellow curves in (b,d) present the third-order phase transition lines, these limit areas with 100\% volume fraction. Black solid   curves point to fifty-fifty volume fraction for two adjacent phases.}
\label{fig4}
\end{figure*}
Obviously, the ground state and $T-n$ phase diagrams of the system described by the effective spin-pseudospin Hamiltonian are determined by the relationship between the values of the parameters characterizing local and nonlocal correlations, the integrals of one- and two-particle transport, as well as spin exchange integrals. A relatively small change in the parameters can radically affect the phase diagram.
 Thus, a small increase in the local correlation parameter $\Delta$ and the transfer integral of composite bosons $t_b$ can lead to suppression of both antiferromagnetic and charge ordering, i.e. AFMI and CO phases in favor of BS and FL phases. It is this situation that is shown in the phase diagram in the Figure\,\ref{fig4}\,a,b, calculated numerically with the parameters shown in the inset to the Figure\,\ref{fig4}a, which presents the results of conventional EFT approach, while the phase diagram in the Figure\,\ref{fig4}b does illustrate the BS-FL interplay with the PS taken into account using the Maxwell's construction.
 Such phase diagrams with no signatures of the long-range AFMI and CO orderings turn out to be typical of cuprates with an ideal, or almost ideal, T$^{\prime}$-structure\,\cite{Naito-2016}, as well as nickelates, the structure of which also lacks apex oxygen\,\cite{Li}.

 For the chosen values of the Hamiltonian parameters, the AFMI phase is energetically unfavorable, but its energy differs little from the energy of the BS and FL phases, so that the slightest 1\% increase in the value of the exchange integral, or a corresponding decrease in the remaining parameters, is sufficient to restore the AFMI phase, albeit in a small region of the phase diagram (see Figures\,\ref{fig4}\,c,d).
 Interestingly, a similar effect of suppression/restoring of the AFMI phase is observed in T$^{\prime}$-cuprates at the slightest change in the concentration of nonstoichiometric apical oxygen\,\cite{Naito-2016}, accompanied by corresponding change of the external potential for CuO$_2$-planes.

\section{Conclusion}
Comparison of the MFA with the results of calculations by the classical and quantum MC methods for the two limits of the spin-pseudospin model shows  the possibilities of MFA for a qualitative and semiquantitative description of phase diagrams and points to the PS as a phenomenon, which is typical for systems with competing order parameters. Effective field theory for complete spin-pseudospin Hamiltonian points to the phase-separated coexistence of main phases AFMI, CO, BS, and FL to be the typical one for the CuO$_2$/NiO$_2$ planes.
Despite the many simplifications associated with the use of MFA, especially for describing 2D planar systems, $nn$- and two-sublattice approximation, the neglect of the doping-dependent nonuniform potential and the screening effects for the Hamiltonian parameters, the use of the simplest version of the Caron-Pratt method for the "real-space" on-site description of the single-particle transport, ..., our model approach is believed to be  a reliable starting point for  description of the 3D phase diagrams in cuprates/nickelates with experimentally observed $T_N$,  $T_c$, and $T^*$ as the critical temperatures for the 3D phase transitions.

\ack
This research was partially supported  by the Ministry of Education and Science of Russian Federation, project No FEUZ-2020-0054.


\end{document}